\documentclass[conference]{IEEEtran}
\IEEEoverridecommandlockouts
\usepackage{cite}
\usepackage{amsmath,amssymb,amsfonts}
\usepackage{algorithmic}
\usepackage{graphicx}
\usepackage{textcomp}
\usepackage{xcolor}
\usepackage{subfigure}
\usepackage{url}
\usepackage{fancyvrb}

\newcommand{\INSTITUTION}{Universidad de Valladolid, Spain}
\newcommand{\Tablon}{Tablon}
\newcommand{\TablonAtTrasgo}{\url{http://trasgo.infor.uva.es/tablon}}

\begin{document}

\title{Carrot and Stick approaches revisited when managing Technical Debt in an educational context\\
\thanks{This work was partially supported by Universidad de 
Valladolid as part of the initiative of grants to innovation in education by means of the projects:
PID2018/2019-38, PID2018/2019-39, PID2019/2020-32, PID2019/2020-, PID2020/2021-.}
}

\author{
\IEEEauthorblockN{Yania Crespo}
\IEEEauthorblockA{
\textit{Universidad de Valladolid}\\
Valladolid, Spain \\
yania@infor.uva.es}

\and

\IEEEauthorblockN{Arturo Gonzalez-Escribano}
\IEEEauthorblockA{
\textit{Universidad de Valladolid}\\
Valladolid, Spain \\
arturo@infor.uva.es}

\and

\IEEEauthorblockN{Mario Piattini}
\IEEEauthorblockA{
\textit{Universidad de Castilla-La Mancha}\\
Ciudad Real, Spain \\
Mario.Piattini@uclm.es}

}

\maketitle

\begin{abstract}
Technical Debt management is an important aspect in the training of
Software Engineering students. In this paper we study the effect of two assessment strategies in an educational context: One based on penalisation, the other based on 
rewards. Both are applied to assignments where the students develop a project focusing
on keeping a low technical debt level, and obtaining a high quality code.
We describe the design, tools and context of the strategies applied.
SonarQube, a tool commonly used in production environments, is used
for measuring the metrics. The penalisation strategy is based on a 
SonarQube quality gate. The reward strategy is based on a contest,
where an automatic judge tool is devised to provide an online leaderboard
with a classification based on the SonarQube metrics.
An empirical study is conducted to determine which of the strategies works better to help the students/trainees keep the Technical Debt low. 
Statistically significant results are obtained in 5 of the 8 analysed metrics, 
showing that the reward strategy works much better. 
The effect size of the executed statistical tests is analysed, resulting in 
medium and large effect size in the majority of the analysed metrics.
\end{abstract}


\section{Introduction}

In 1992, Ward Cunningham coined the term \emph{``Technical Debt''} \cite{cunningham1993wycash}. If there exists a high technical debt in a software project, development and maintenance will become too slow and laborious. According to Cunningham, this is similar to paying interest on a financial debt. 
As stated in~\cite{avgeriou_et_al:DR:2016:6693}: 
``While the conceptual roots of technical debt imply an idealized, deliberate decision-making process and rework strategy as needed, we now understand that technical debt is often incurred unintentionally and catches software developers by surprise \ldots\ Managing Technical Debt includes recognizing, analyzing, monitoring, and measuring it.''

For all this, a very important aspect in the current training of the Software Engineering student, as a future professional of software development, is the preparation in terms of quality software development and Technical Debt (TD) management \cite{Allman2012}. From the developer's point of view, the emphasis would be on keeping TD under control at development time and reducing it as much as possible at maintenance time. Five reasons why TD should be included in the Software Engineering curriculum are presented in \cite{FiveReasons2015}. The authors claim that TD should be treated with the same relevance as Requirements Engineering, Software Design and Architecture, and Software Testing.

An empirical study on how managers communicate and reward/encourage/penalise/force developers in today's software industry to keep TD as low as possible was presented in \cite{CarrotStick2020}. The study was surveyed 258 participants and interviewed 32 developers in the industry. The authors concluded that developers are not commonly rewarded, penalised, or
forced to keep the level of TD down in their companies. $60\%$ of the 
respondents stated that they are, to some extent, encouraged to keep
the level of TD down. The authors also stated that a TD
management strategy based on encouraging activities is described
as having a significant impact on software engineers' attitudes and
behaviour when addressing TD.

In an educational environment, instructors usually \textbf{encourage} students to keep the TD as low as possible in their assignments. 
In this work:
\begin{itemize}
    \item We study and analyse the impact on learning outcomes related to the TD using two different assessment strategies.
    The first is an approach based on penalties in grading. 
    The second is based on rewards using gamification techniques. 

    \item We present the design of the gamified reward strategy based on a contest. We discuss the tools used to implement it, with an online score leaderboard, in a Software Engineering course.
    
    \item We present an empirical study comparing measures obtained from students' assignments that work using one or the other strategy. 
\end{itemize}

The rest of the paper is organised according to the following structure. Section~\ref{sec:relatedwork} presents related work. Sections~\ref{sec:penalising}
and~\ref{sec:rewarding} explain both strategies, penalising and rewarding, respectively, in the educational context they were developed. The experimental design governing the conducted study is presented in Section~\ref{sec:Methodology}. The results obtained are presented in Section~\ref{sec:results}. The threads to the validity of the study are discussed in Section~\ref{sec:threads}. Finally, Section~\ref{sec:conclusions} concludes. 

\section{Related Work}
\label{sec:relatedwork}

As mentioned before, the authors of \cite{CarrotStick2020} 
analyse the reward/encourage/penalise/force strategies for TD management in companies. In an educational context, the encouragement is always present in the teachings of instructors. The impact of encouraging is very difficult to measure. Strategies using force, in the terms assumed in  \cite{CarrotStick2020}, such as forcing to keep the level of TD down by not allowing deployment or delivery, would imply, in an educational context, preventing assignment delivery or grading. This strategy is not usually put into practice, because it prevents the evaluation of student performance and learning, and can lead to a total lack of motivation. Our study  comparatively analyses the impact of penalty and reward strategies on the quality of the code and the level of TD of the students' projects. To the best of our knowledge there is no precedent of studying  penalisation as opposed to rewarding strategies in the context of learning and training development of quality code and TD control.

Rewarding strategies are closely related to gamification and serious games. Software Engineering and Programming subjects are the natural space for training in code quality and TD. Several systematic literature reviews have been conducted regarding gamification in Software Engineering, such as \cite{Pedreira2015}, \cite{Alhammad2018} and \cite{GamificationInSoftwareEngineering2021}. The last of these three focuses on non-educational contexts of software engineering 
activities. It concludes that some
 major benefits are achieved with the use of gamification, such as
engagement and motivation to
 perform the activities, as well as encouragement for code review tasks, among others. 
Another finding is that  despite the existence of several gamification elements, the most commonly used in the investigated context are points and leaderboards. 

Antecedents on gamification regarding TD can be found in the well known games ``Hard choices'' \cite{HardChoiceGame2011, HardChoiceGame2017whitepaper} and ``Dice of Debt'' \cite{DiceOfDebt2016}. Both are board games that encourage discussion and decision making when managing technical debts, understanding the risks and costs. In \cite{HardChoiceGameExperience2014}, an experience of conducting the Hard Choice game in a classroom was analysed. 
Pre- and post- tests are applied, showing that after conducting the board game the scores obtained in the post-test were significantly higher. The results of a questionnaire also show that the students perceived the instruction using the board game as engaging. 

Serious games for teaching and training code reviews 
are presented in \cite{SeriousGameDecisionMakingCodeReview2015} and \cite{SeriousGameTeachingCodeReviews2020}.
In \cite{SeriousGameDecisionMakingCodeReview2015}, the player 
plays the role of a project
manager and a scoring system as a factor of time, budget, quality
and TD is defined. Pre-game and post-game questionnaires were used to evaluate the proposed game with the participation of 17 students. Although TD is important in scoring, it is not appreciated from the evaluation that the game leads to learning related to this aspect. On the other hand, the game presented in \cite{SeriousGameTeachingCodeReviews2020} is based on code segments, the higher the game level the larger the segment, with numerous code defects planted in them. The player has to find all of them, along with their classifications. Training code defects identification is closely related to TD knowledge. An evaluation was carried out in two different courses, with 52 and 80 students, with a survey after the game sessions. The students were asked to evaluate the usefulness and their enjoyment of the game components on a 5 point Likert scale. No report on the total of students answering the survey was presented nor any evaluation or study of student learning or outcomes. None of the two mention which tool(s) (if any) was used to score TD in the first case or to prepare the code base with defects in the second case. 

In \cite{SeriousRefactoringGame2019}, a serious game design for training in refactoring to improve software quality was proposed. The theoretical design is based on existing tools for assessing the quality of systems such as SonarQube, NDepend or JArchitect. The game mechanics is based on: (1) search for and select refactoring candidates; (2) plan and perform source-code modification; and (3) check the feedback/result in terms of the system’s behaviour and quality. The behaviour should be checked with test frameworks  in terms of
runtime regression tests to assure a source-code transformation that preserves the behaviour. The score is based on TD measured by a quality analyser, such as SonarQube, after the tests have run successfully.
The authors define three game modes: single-player, parallel multi-player, and alternating multi-player modes. A theoretical scenario of game application is described. In contrast with our study, the proposal is theoretical and no real experience is reported. The rewarding strategy we defined is based on gamification, while the scoring is based on a combination of measures obtained from SonarQube, as described in Section~\ref{sec:rewarding}.

The benefits of gamification in introductory programming courses are dealt with in \cite{BenefitsOfGamification2018} by conducting a quasi-experiment to obtain empirical evidence on the improvement of students’ learning performance when using a gamified compiler in comparison to a non-gamified compiler. The authors found statistical significance in favour of there being a positive effect on the learning performance of those students who used the gamified platform.

The authors of~\cite{GamificationMotivateStudentsQualityCoding2019} 
go further, applying gamification to motivate students to improve
the quality of the code in programming assignments in an intermediate programming course. The authors introduce scoring based on measuring Cyclomatic Complexity in an online judge. 
The students answer a programming problem by submitting
code to the online judge system. The system measures the Cyclomatic Complexity of the submitted code, scores based on this measure, and shows the score and the leaderboard
to the student. The students can resubmit his/her code zero or more times, and the system records the smallest
Cyclomatic Complexity based score. The proposal was evaluated over 6 weeks with the participants in a C programming course. An experiment was conducted randomly dividing 35 students into two subgroups, and also dividing the 6 week frame into two periods. Meta-analysis to the cyclomatic complexity measures of the submissions for each programming problem was applied. The results indicate that students wrote code with a lower cyclomatic complexity under gamification
conditions.

More closely related to our work is the proposal in \cite{GamifyingStaticAnalysis2018} for gamifying static analysis. The authors presented a theoretical proposal (an initial UI prototype of a gamified static analysis tool). An early evaluation of the proposed UI was conducted with eight researchers who have knowledge of how static analysis
tools function. The evaluation consisted of a cognitive walkthrough of the prototype having to perform (theoretically) 23 tasks grouped in five
themes. The participants find the UI for gamified static analysis tool useful to complete their tasks as well as engaging.

Recently, in
\cite{CodeQualityInOnlineProgrammingJudge2020}, an online judge system is enriched in a similar way to \cite{GamificationMotivateStudentsQualityCoding2019}, but using 
SonarQube to provide code quality detection in the online judge system. There are no penalties or rewards, nor any gamification approach. The goal is to enrich the usual function of the online judge, assessing the correctness of the problem solution based on the execution of test sets, with the quality issues detected by SonarQube. The authors collect data across different introductory programming courses to obtain a ranking of the most frequent issues in Python code. The authors recognise that the students pay lots of attention to solving the
assignment problems to generate correct answers, but omit the
code quality issues when they deliver program codes. The
instructor should emphasise (encourage) the concepts of code quality in programming education.

In short, compared to the related work, this paper studies the effects of penalisation and reward strategies \textbf{in an educational context}, while comparing the \textbf{students' outcomes in terms of metrics measured on their projects}. The reward is based on points obtained during a contest to keep the technical debt level low and to increase the quality of the developed code. Unlike other works previously discussed, the \textbf{contest has been fully implemented}, it is not a theoretical proposal, and \textbf{several metrics related to the technical debt and  quality of the developed code are integrated into the scoring and tie resolution}. The design of the execution environment prioritises the use of \textbf{SonarQube as it is commonly used in production}, but adding the game elements based on scores, ranking, and leaderboards. A study is conducted based on a \textbf{rigorous experimental design}, 
analyzing the \emph{effect size} of the executed statistical tests.   

\begin{figure*}[htbp]
\centerline{\includegraphics[width=0.8\textwidth]{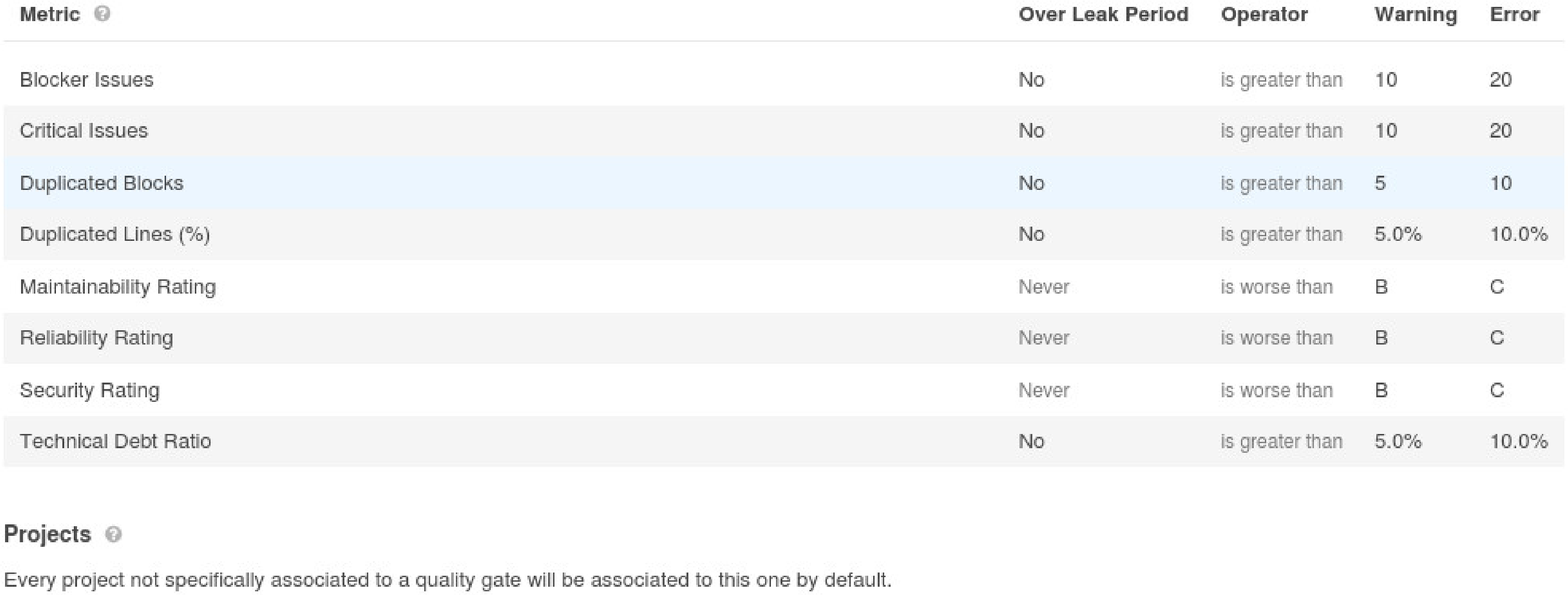}}
\caption{Quality gate defined and set by default in the SonarQube server.}
\label{fig:quality_gate}
\end{figure*}

\section{Penalising Technical Debt}
\label{sec:penalising}

This study is done in the context of
a \emph{Software Architecture and Design}
course of a major in \emph{Software Engineering} at \INSTITUTION{}.
In this course, the students develop a software project over 8/9 weeks.
They work in teams of 3 or 4 members. The assignment consists 
of performing the design and implementation of a system given the 
requirements and analysis documentation. 
They should document the architectural and design decisions with UML diagrams and perform the detailed design and programming of 4 of the use cases described in 
the requirements and analysis documents and models. 
The instructors always encourage them to develop a high-quality code.
Code smells, TD, and other aspects related to security, reliability, maintainability, etc., are explained in the classroom. 

Since the academic year 2017-2018, students have a SonarQube server installed at the institution. This server has been properly configured so that they can 
submit their projects to analyse them
throughout the entire development of their assignment. 
During the academic year 2017-2018, the instructors applied the strategy of simply encouraging them to use the server. At the end of the course,
it was found that the students had not performed any analysis with the SonarQube server.

With the aim of training students in recognising, analysing, 
and measuring TD during the academic year 2018-2019, 
a penalty-based approach was applied. 
The assignment statement included a penalisation on the project grade based
on the use of the server:
``The quality and security of the application code will be analysed 
by SonarQube. Quality and security problems detected by this tool 
will lead to a penalty of up to $10\%$ in the mark assessing the 
application code.''

To implement it, 
a Quality Gate is defined in the SonarQube server. 
This Quality Gate is assigned by default to all the students' projects 
in the course. Fig.~\ref{fig:quality_gate} shows the definition of 
the Quality Gate in the SonarQube server.
The project is not penalised when it passes the Quality Gate 
(green mark: Passed). A $10\%$ penalty is applied when the project does 
not pass the Quality Gate (red mark: Failed), and a $5\%$ penalty when 
the project is warned (yellow mark).

The quality gate was set by analyzing the code quality of the projects developed by students in previous courses, concluding that improving, in general, would require this as a minimum.

At the end of the academic year, the results indicate that all projects, except one, 
were penalised. To sum up, the enrolled students (48) presented a total of 13 projects. Six of them were penalised with $10\%$, six with $5\%$, 
and only one was not penalised. 

\textbf{The study} presented here \textbf{involves 11} of these 13 projects,
discarding the first versions of two projects that were allowed to be resubmitted as new ones in the case of students that did not pass the first evaluation 
of the course.

\section{The rewarding strategy}
\label{sec:rewarding}

In the academic year 2019-2020, a different reward-based approach 
was implemented. The reward is obtained by participating in a contest. 
A gamification strategy based on a leaderboard is applied with prizes in points 
that are added to the assignment grade.
An automatic online judge tool is introduced to provide immediate feedback
to the students and encourage the competition. The description of the
contest environment is presented in section~\ref{sec:environment} and the details on the contest rules are given in section~\ref{sec:Rules}.

\subsection{The contest environment}
\label{sec:environment}

\begin{figure}[htbp]
\centerline{\includegraphics[width=\linewidth]{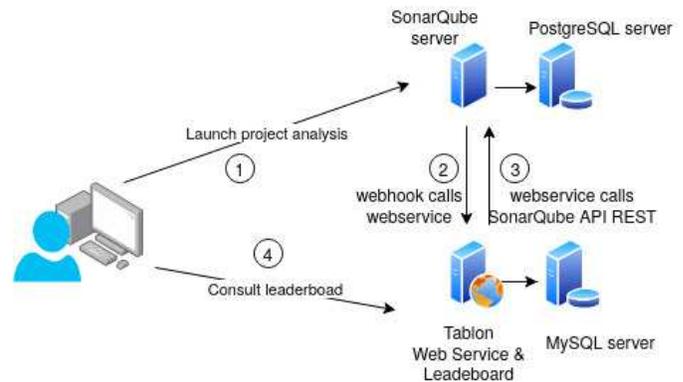}}
\caption{Contest implementation environment}
\label{fig:ContestEnvironment}
\end{figure}

To carry out the contest in a motivating way for the students,
an online judge tool and leaderboard system is used. 
Fig.~\ref{fig:ContestEnvironment} shows the implementation 
of the contest environment. 
It is made up 4 servers on premises: (a) SonarQube server, 
(b) PostgreSQL server for SonarQube database (c) Web server with a web 
leaderboard application (\Tablon), and (d) MySQL server used by
\Tablon{} to store data about each submission, scores and ranking information.

The \Tablon{} leaderboard application 
was originally developed to introduce
gamification in the context of Parallel and High Performance Computing 
courses~\cite{FresnoEt17,GonzalezEt19}.
It was focused on compiling and executing codes 
in back-end systems, measuring their performance. 
The only ranking criteria was a single parameter, execution time. 
For this experience, the \Tablon{} application has been improved to also
support a new type of queue and leaderboard.
The new type of queue supports submissions containing data for 
several metrics or precalculated scores.
The new leaderboard type can 
compute a final score from the metrics received, and use the
original metric values as tie-breaking criteria in case it 
is needed (see contest rules in section~\ref{sec:Rules}.
\Tablon{} stores information about each submission received 
and updates the leaderboard information on the fly. It has a web service
front-end that can show the results of each submission, the submissions list, 
a leaderboard page with the current ranking, and several items of statistical 
information about users, quotas, submissions by time range, 
or a global view of the positions of the users on the leaderboard
day by day. 
These features are designed to 
encourage competition,
offering the students an on-line service
with immediate feed-back on each submission and historical information
about their evolution. 
\Tablon{} is a python program that uses external scripts to interact with
other programs. It is an open source project
(check \TablonAtTrasgo).

To connect the different services, 
a webhook was configured in the SonarQube server so that each time a team 
launched an analysis, a web service was invoked. The web service collect all 
the necessary measures by calling the SonarQube's API REST. 
This web service also calculates the partial score and supplies all the data 
to the \Tablon{} tool. As mentioned before, \Tablon{} is responsible for recording the measures, maintaining the leaderboard, and producing all the statistics.

\begin{figure*}[htbp]
\centerline{\includegraphics[width=0.8\textwidth]{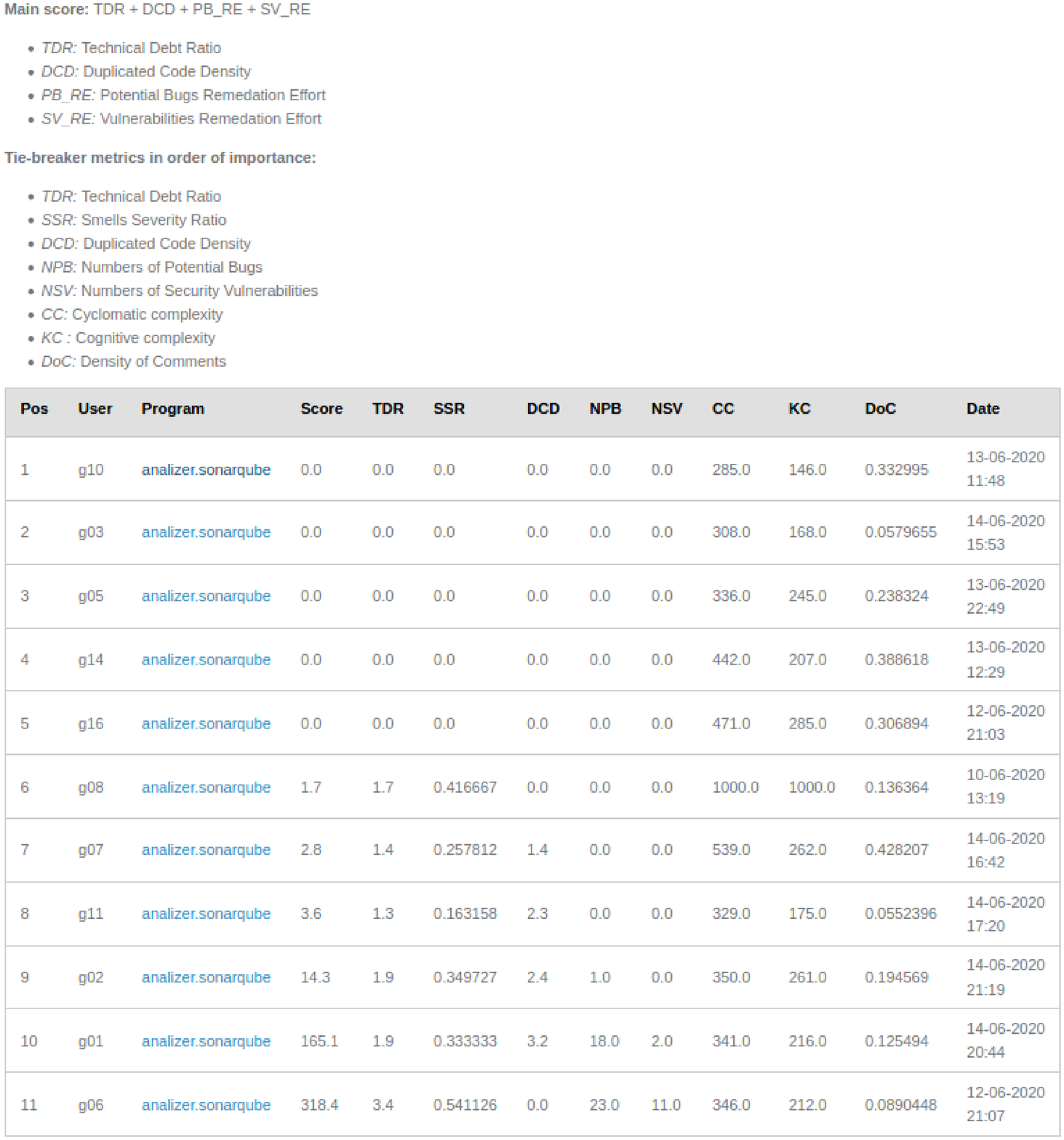}}
\caption{Example of the leaderboard web page showing the final classification with the best submission of each user.}
\label{fig:leaderboard}
\end{figure*}

As Fig.~\ref{fig:ContestEnvironment} shows:
\begin{enumerate}
    \item The students launch the SonarQube analysis of their projects 
		using the SonarScanner.  
    \item A webhook in the SonarQube server is activated and calls the web 
			service identifying the project.
    \item The web service calls the SonarQube's API REST to collect measures 
			of the metrics required, scores the project and 
			submits the information to the \Tablon{} leaderboard application.
			The submission information, new measures and score are recorded.
			The leaderboard and statistics are updated.
    \item The students can see the results of the submission and the
			updated leaderboard by accessing the \Tablon{} web application,
			as shown in Fig.~\ref{fig:leaderboard}.
\end{enumerate}

All this process is repeated over the development time until the delivery deadline is reached. 

\subsection{Contest rules}
\label{sec:Rules}

The contest rules are defined in this section. They were given to the students in the assignment statement document. There are two prerequisites to qualify for the ranking to guarantee a certain quality level and to enforce fairness.

To qualify for the ranking, the following is required: 
\begin{itemize}
    \item 
To pass the QualityGate (green mark: Passed) defined in the SonarQube server. This Quality Gate is the same as the one in the previous academic year (see Fig.~\ref{fig:quality_gate}).
    \item 
To have implemented all the functionality specified in the use cases given in the assignment statement.
\end{itemize}

Once qualified for the ranking, each team is scored.
The lower the score, the higher the position in the ranking.
The score is calculated as:
$TDR + DCD + PB\_RE + SV\_RE$,
where:\newline
$TDR$ is the accumulated technical debt ratio,\newline
$DCD$ is the density of duplicated code,\newline
$PB\_RE$ is a rate of the potential bugs remediation effort, and\newline
$SV\_RE$ is a rate of the vulnerabilities remediation effort.

As Fig~\ref{fig:leaderboard} shows, the ties are solved applying the following criteria, in this order:
lowest technical debt ratio,
lowest code smell severity defined as (blocking+major)/violations, 
lowest duplicated code density,
lowest potential bugs,
lowest vulnerabilities,
lowest cyclomatic complexity,
lowest cognitive complexity,
and finally, the one with the highest density of JavaDoc comments (although not evaluated in this assignment, students are aware of their importance for maintainability).

The reward is set as an extra bonus in the grade of the assignment, for 
the first three teams classified in the ranking. No ties are allowed. 
The team in first place obtains $0.9$; second place: $0.6$; third place: $0.3$. 
If the total grade after adding the bonus is higher than the maximum grade
of the assignment, it is truncated to the maximum.

In that academic year, there were 55 students enrolled, organized in 17 teams.
14 projects were submitted. 
11 of them competed in the contest and the remaining 3 did not 
meet the prerequisites 
to qualify for the ranking.
However, \textbf{in the study} presented here \textbf{13 projects} out of these 14 were included because they participated in the experience using
the reward strategy. 
A project is excluded because it only presented the design and did not make the implementation. Thus, there is no code to analyze with SonarQube.

Among the 11 teams qualified for the ranking, early during the contest, there 
was a five-fold tie in the first position, with 0 points.
The competition 
continued until the end of the contest, with these teams
competing to improve their codes according to the criteria established 
to solve the ties.

Fig.~\ref{fig:leaderboard} shows the final leaderboard classification.
The team occupying the first position in the ranking was not awarded the bonus 
at the end, because they did not
fulfilled the prerequisite ''to have implemented all the functionality specified in the use cases given in the assignment statement''. 
In \url{http://tablon-ds.infor.uva.es/stats/leaderboard?lb=concurso}, different data about the 
competition are shown.

\section{Experimental design}
\label{sec:Methodology}

Having explained the context of the two experiences based on penalising and rewarding through gamification, this section presents the empirical study conducted to comparatively analyse both approaches.

\subsection{Goal}

The goal of the experiment is to compare the Penalising TD strategy with the Rewarding Strategy for the purpose of evaluating TD control and code quality in the context of a Software Architecture and Design course.

\subsection{Research questions}

\begin{description}
\item[RQ1:] Are better results obtained in terms of technical debt control and code quality when a reward strategy is applied or when a penalty strategy is applied? 
\item[RQ2:] If better results are obtained with one strategy, can the improvement be considered significant?
\end{description}

\subsection{Subjects}

Subjects were students following the Software Architecture and Design course of a major in Software Engineering at the \INSTITUTION{} in the academic years 2018-2019 and 2019-2020.

\subsection{Apparatus and experimental tasks}

In both years, students and teachers worked with the same SonarQube instance. All the metrics that participate in the definition of the contest and that students can visualise when analysing their projects with SonarQube will be taken as a reference for the study. These same metrics influenced the penalty in one year and have their role in the contest-based reward in the other year.

The instructors run the analyses with SonarQube of all the projects participating in the study to ensure that the analyses are carried out under the same conditions. In this course, we do not work with maven/ant/gradle/... project automation, so the analysis of SonarQube is launched using the SonarScanner and the configuration of the property file 
\verb|sonar-project.properties|.

For each project, the property file must be similar to:
\begin{Verbatim}[fontsize=\small]
sonar.host.url=https://sonarqube.anonymous.url
sonar.login=??????????????????????????
# must be unique in a given SonarQube instance
sonar.projectKey=sd-19-20-??
sonar.projectVersion=1.0
sonar.sources=src/main/java/
sonar.java.binaries=target/classes
sonar.sourceEncoding=UTF-8
sonar.exclusions = **/View*.java
\end{Verbatim}

 Each project has a unique identifier given by the academic year 18-19 or 19-20 and two digits
as the team number in the academic year. The \verb|sonar.login| property is set with the security token used to analyse all the projects, the instructor's security token. The property \verb|sonar.exclusions| is set with a pattern of file names representing the Views of the application. The projects are developed with Netbeans, and the views are automatically generated with the interface designer. The code automatically generated by Netbeans for the views cannot be modified by the students, so it does not participate in the analysis.

The study is conducted using a SonarQube server on premises version 7.3. The analysis of all the projects participating in the study are executed using the SonarScanner and the \verb|sonar-project.properties| files configured for each project with the corresponding key and exclusions. 

The data collected for the study have been read once at the end of both experiences. To guarantee, as mentioned before, that all data are collected the same way using the same criteria with the same SonarQube server. Intermediary measures cannot be obtained because not all teams used version control software. Nevertheless, in the case of the rewarding strategy, the evolution of measurements can be obtained because they are stored in the leaderboard's database.

Once the participating projects are hosted on the SonarQube server, SonarQube's  API REST services are used to obtain the measures of the direct metrics. The fundamental service here is \verb|api/measures/component|. 
The tool R with RStudio is used to automate this process. Once the direct metrics are obtained for each project, the derived metrics are calculated. The data is organized, tagged and cleaned using functions provided by the \verb|tidyverse| packages. Finally, the dataset is saved as a .csv file. This dataset can be obtained from
\url{https://github.com/yania/TechnicalDebt2021}. 
From this .csv file, the rest of the study execution is performed as presented below. The R script and results are also available at the given github repo.

\subsection{Variables}

The \textbf{independent variable} strategy is defined, which takes two possible values for penalising and rewarding (factor type, nominal scale). 
All the projects of the academic year 18-19 are labelled with the penalising value for the strategy variable.
All projects in the academic year 19-20 are labelled with the rewarding value for the strategy variable.

The \textbf{dependent variables} are the following, obtained directly with SonarQube: 
\begin{itemize}
\item SQALE Technical Debt ratio, the smaller the better
\item Duplicated lines density, the smaller the better
\item Comment Density, the larger the better, as long as the principles of good javadoc comments and the principle ``if you feel the need to write a comment, write a function'' are observed.
\end{itemize}
and the derived metrics calculated from the metrics directly obtained from SonarQube: 
\begin{itemize}
\item Smell density = code\_smells/ncloc, the smaller the better
\item Security rate = security\_remediation\_effort/ncloc, the smaller the better
\item Reliability rate = reliability\_remediation\_effort/ncloc, the smaller the better
\item Cognitive complexity rate = cognitive\_complexity/ncloc, the smaller the better
\item Cyclomatic complexity rate = (complexity-functions)/ncloc, the smaller the better
\end{itemize}

\subsection{Statistical hypothesis testing}

First, a descriptive study using box-plots will allow the behaviour of the dependent variables in the two groups of the study (penalising and rewarding) to be visualised. 

Subsequently, the Normality of the dependent variables is studied, in order to use, if possible, the most powerful test to compare two independent samples, such as the observations of each dependent variable in the two groups of the study.

To answer RQ1, the \textit{t-student} test for two independent samples will be used if Normality is accepted, otherwise the \textit{Wilcoxon} test for two independent samples will be used. 

Once Normality is not accepted, to answer RQ1, the Wilcoxon test for two independent samples will be applied to determine if the measures obtained with each strategy are different. The Wilcoxon test is a non-parametric alternative to the t-test for comparing two means. It is particularly recommended in a situation where the data are not normally distributed. We use the function 
\verb|coin::wilcox_test()| in R  because it provides the $Z$ statistic, which is used to calculate the effect size. The function \verb|coin::wilcox_test()| implements the Asymptotic Wilcoxon-Mann-Whitney Test.

The null hypothesis of equality of the distribution of $Y$ (values of each dependent variable in the study) in the groups defined by the independent variable strategy (as factor: penalising, rewarding) is tested against shift alternatives. The ``two-sided'' null hypothesis is $H_0: \mu = 0$, where $\mu = Y_1 - Y_2$ and $Y_s$ is the median of the responses in the group determined by each value of the independent variable strategy. When the alternative = ``less'', the null hypothesis is $H_0: \mu >= 0$. When the alternative = 
``greater'', the null hypothesis is $H_0: \mu <= 0$. 

Our goal is to check not just that both groups present a different behaviour (``two-sided'' case) but to check whether the behaviour is better. The meaning of better depends on each metric. As stated in Section~\ref{sec:Methodology}, the measures of the metrics in the study are best when the value is smaller, except for the metric Comment density, considered best with a greater value. As the comparison is executed by comparing `penalising' with `rewarding' we use the ``greater'' alternative with all the metrics, except in the case of Comment density. In this case, the ``less'' alternative is used. Rejecting $H_0$ means accepting that penalising implies that the measures of the metrics are greater than (worse than) the measures in the case of rewarding; except in the case of Comment density, where rejecting $H_0$ means accepting that penalising implies values of Comment density less than (worse than) the values obtained in the case of rewarding.  

To answer RQ2, the effect size will be calculated and Cohen's classification will be applied \cite{Cohen88}.  The interpretation of the effect size statistic will be based on this classification to answer RQ2 as follows: 
 $0.10 - < 0.3$ (small effect), $0.30 - < 0.5$ (moderate effect) and $>= 0.5$ (large effect).

\section{Results}
\label{sec:results}

\subsection{Overall results}
\label{sec:overallresults}

\begin{figure*}[htbp]
    \begin{center}
    \subfigure[SQALE Technical Debt ratio]{%
            \label{fig:boxplot_sqale_ratio}
            \includegraphics[width=0.2\textwidth,trim={0 0.3cm 0 0.8cm}]{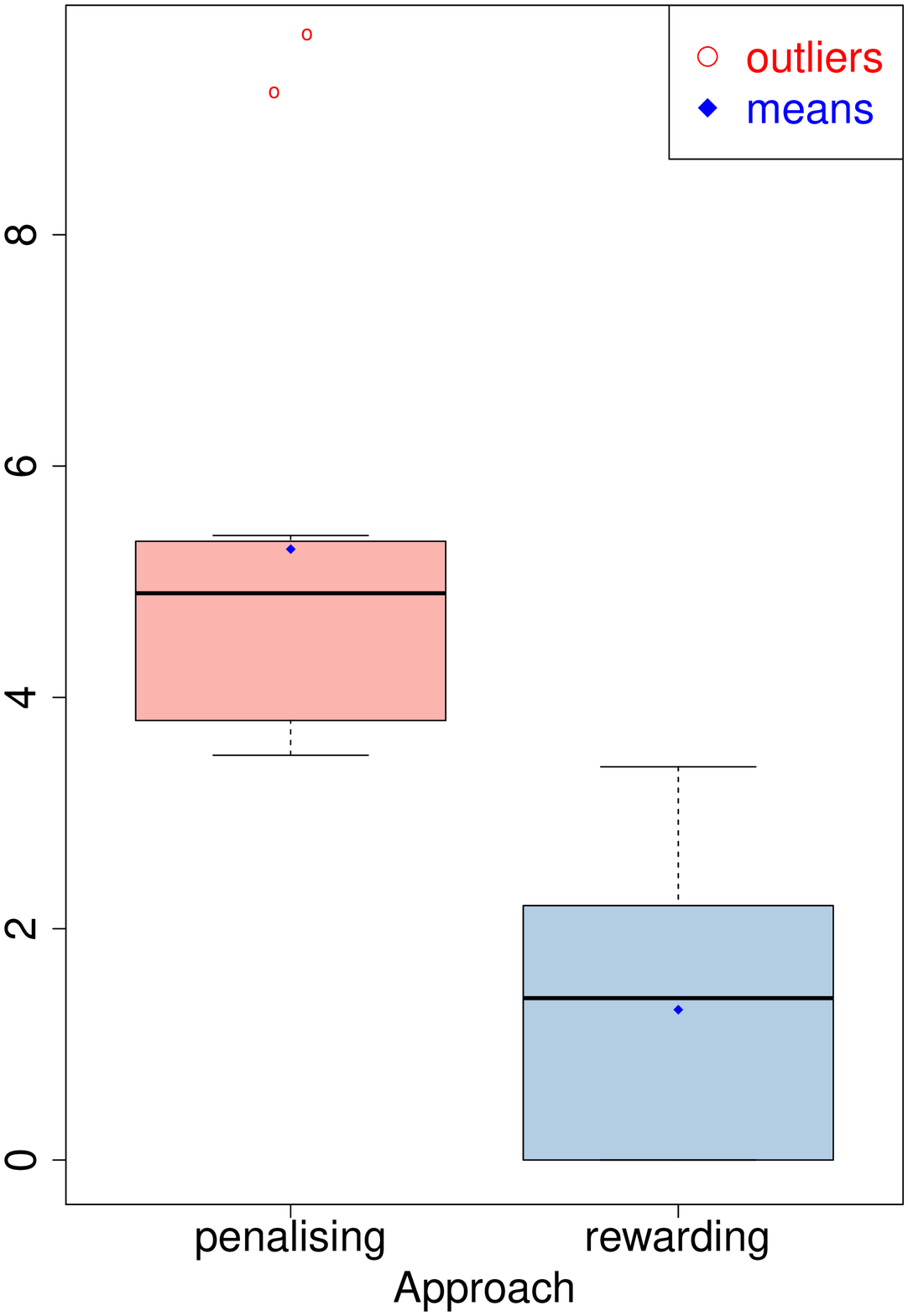}
    }%
    \subfigure[Smells density]{%
            \label{fig:boxplot_smellsdensity}
            \includegraphics[width=0.2\textwidth,trim={0 0.3cm 0 0.8cm}]{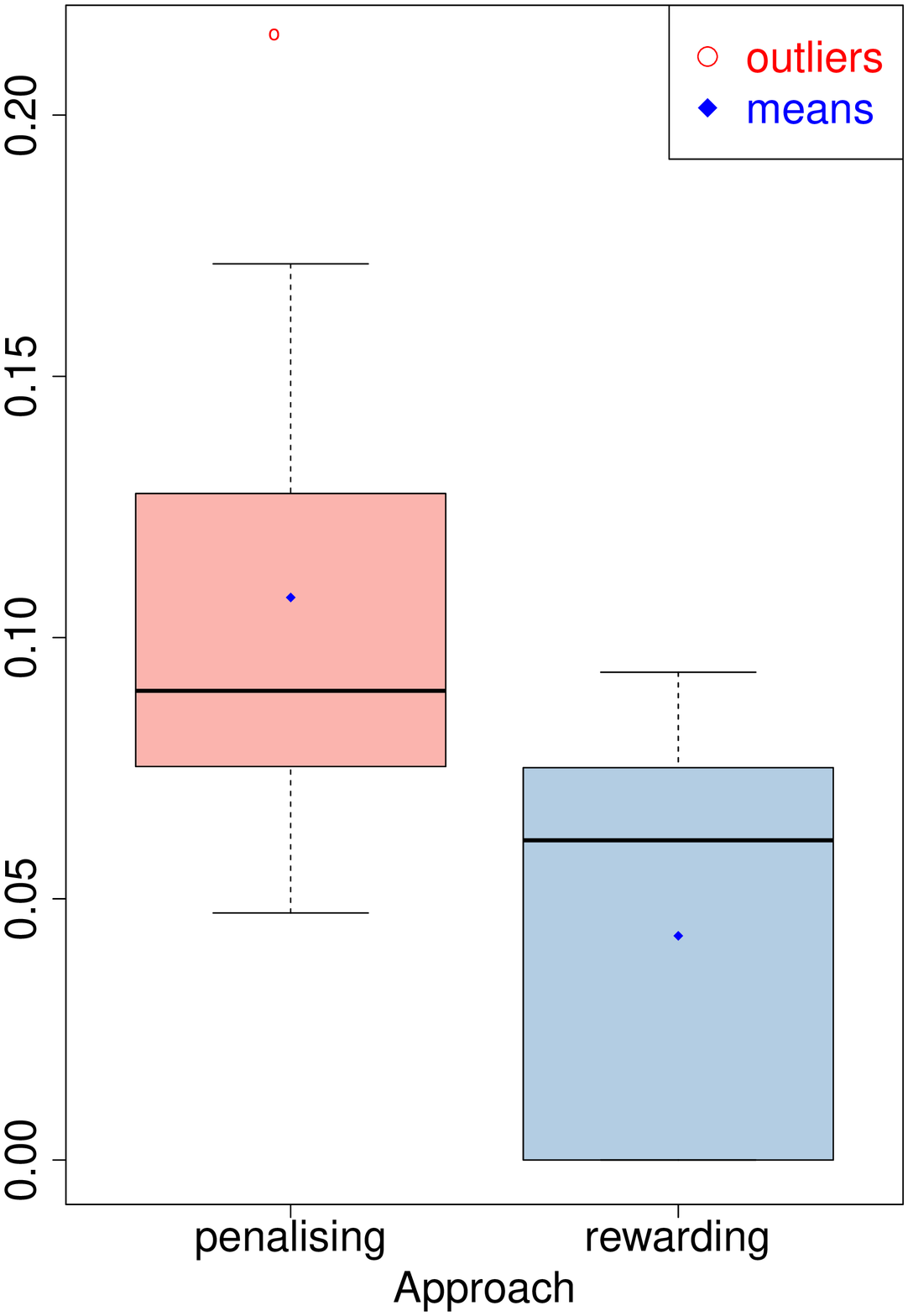}
    }%
    \subfigure[Duplicated lines density]{%
            \label{fig:boxplot_duplicatedlinesdensity}
            \includegraphics[width=0.2\textwidth,trim={0 0.3cm 0 0.8cm}]{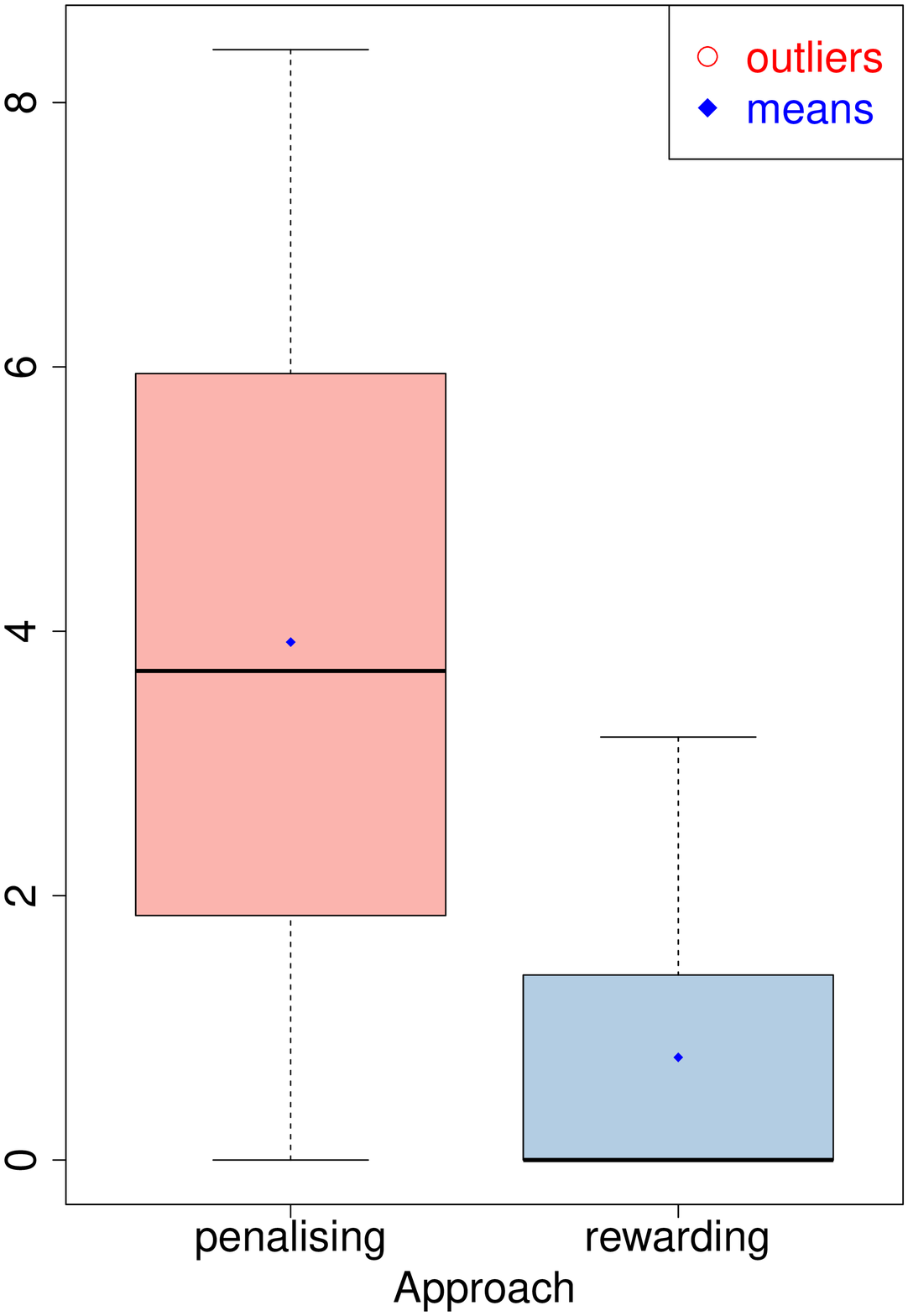}
    }%
    \subfigure[Comment density]{%
            \label{fig:boxplot_commentdensity}
            \includegraphics[width=0.2\textwidth,trim={0 0.3cm 0 0.8cm}]{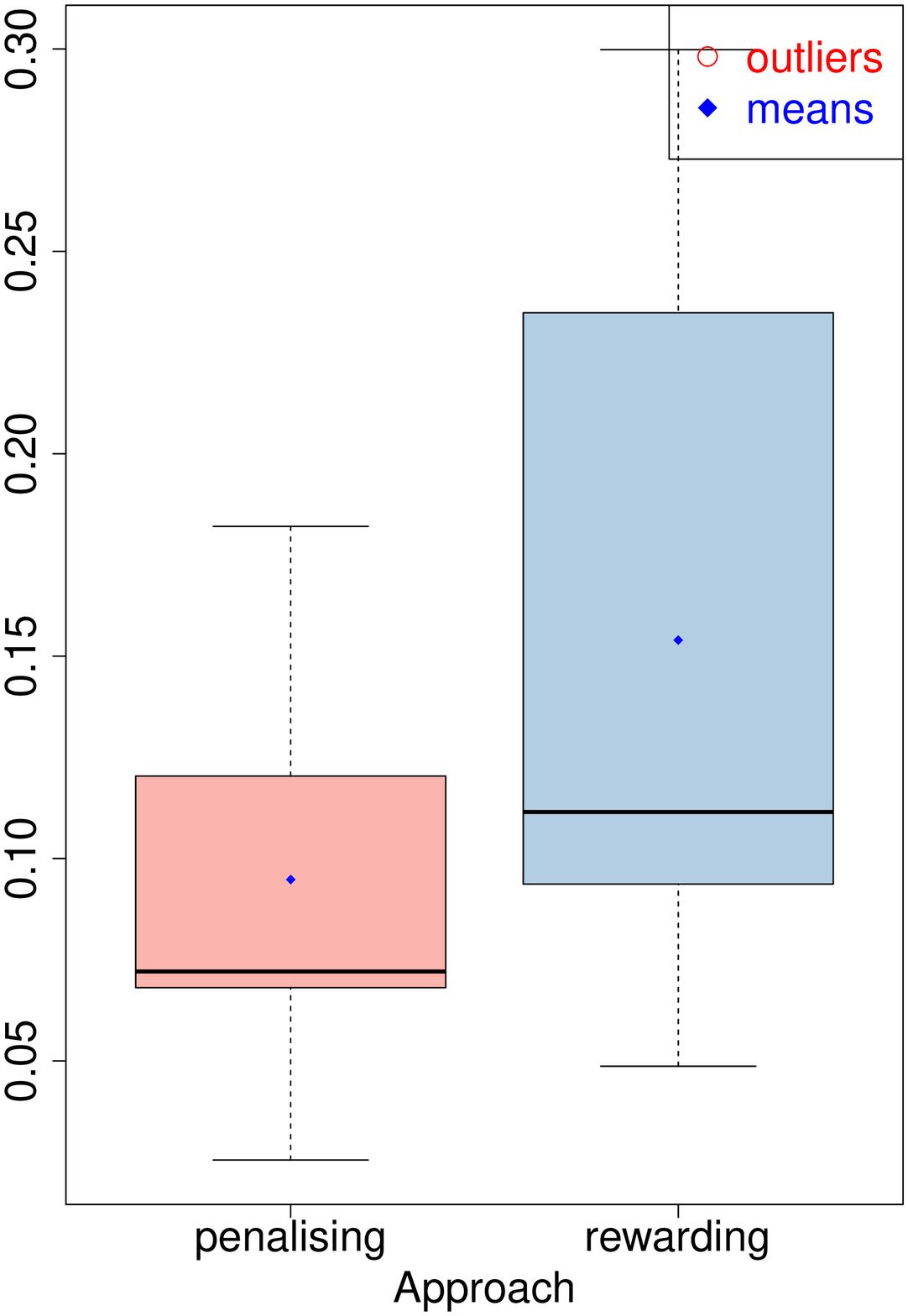}
    }%
    \\
    \subfigure[Security rate]{%
            \label{fig:boxplot_securityrate}
            \includegraphics[width=0.2\textwidth,trim={0 0.3cm 0 0.8cm}]{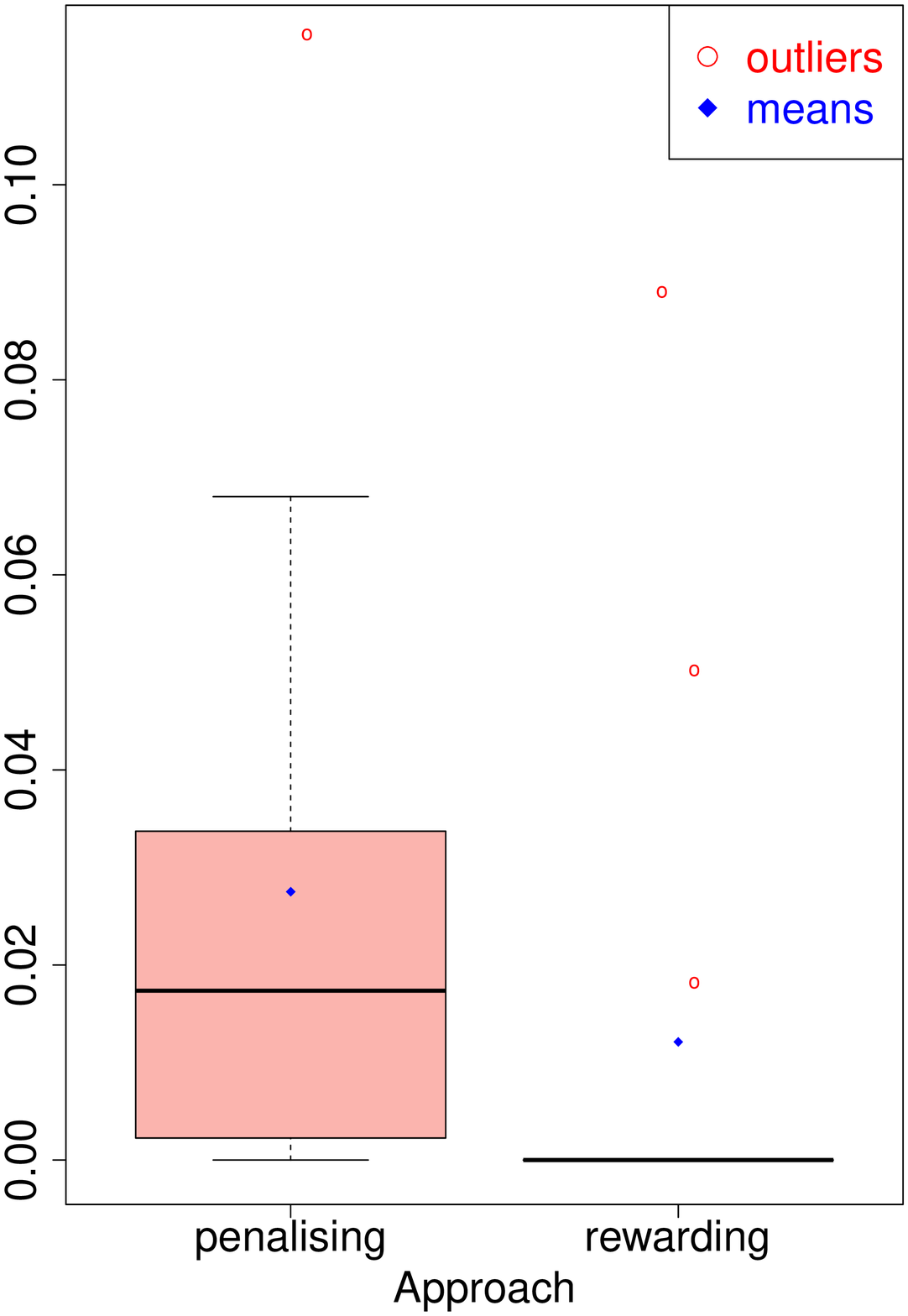}
    }%
    \subfigure[Reliability rate]{%
            \label{fig:boxplot_reliabilityrate}
            \includegraphics[width=0.2\textwidth,trim={0 0.3cm 0 0.8cm}]{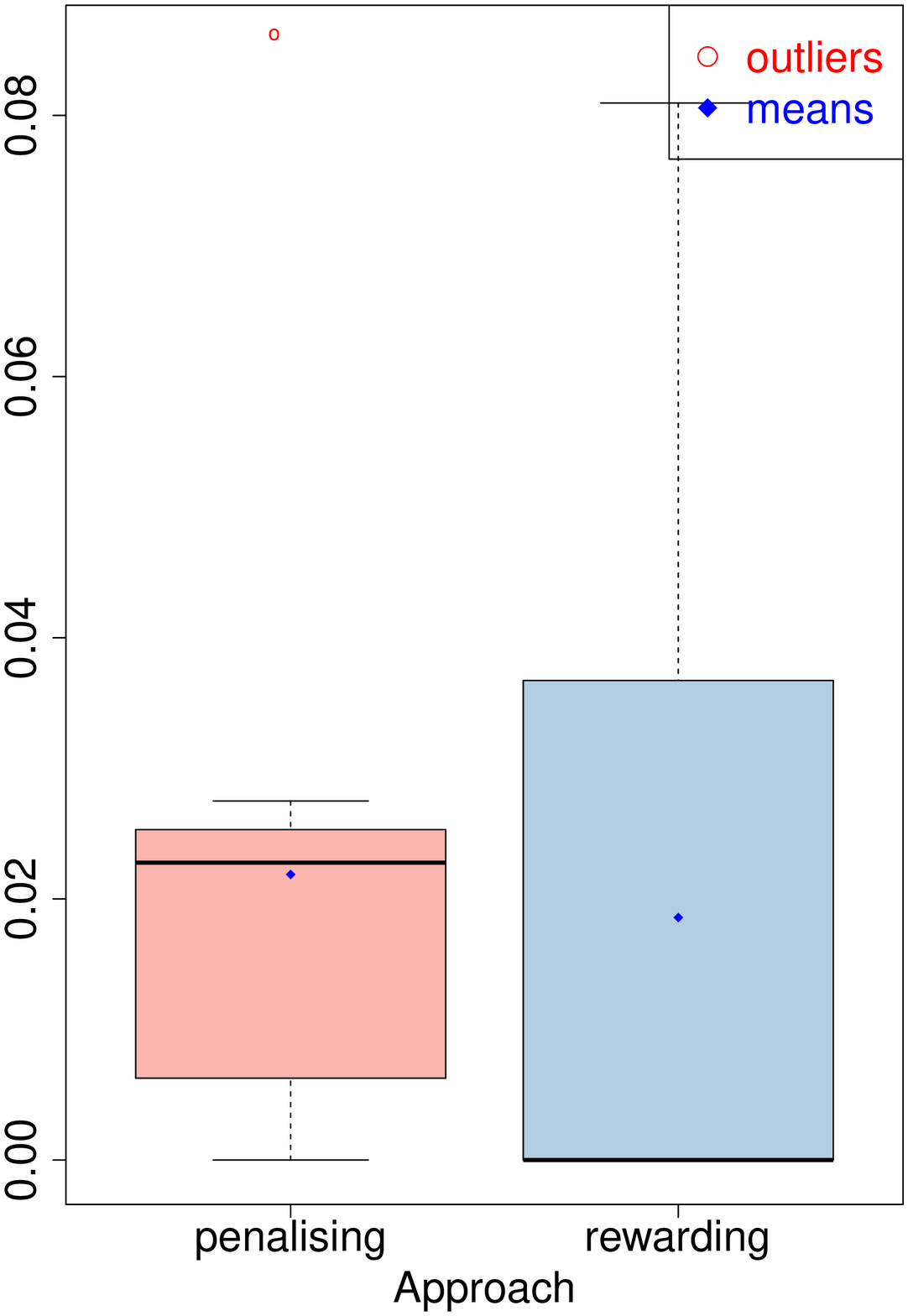}
    }%
    \subfigure[Cognitive complexity rate]{%
            \label{fig:boxplot_cognitivecomplexityrate}
            \includegraphics[width=0.2\textwidth,trim={0 0.3cm 0 0.8cm}]{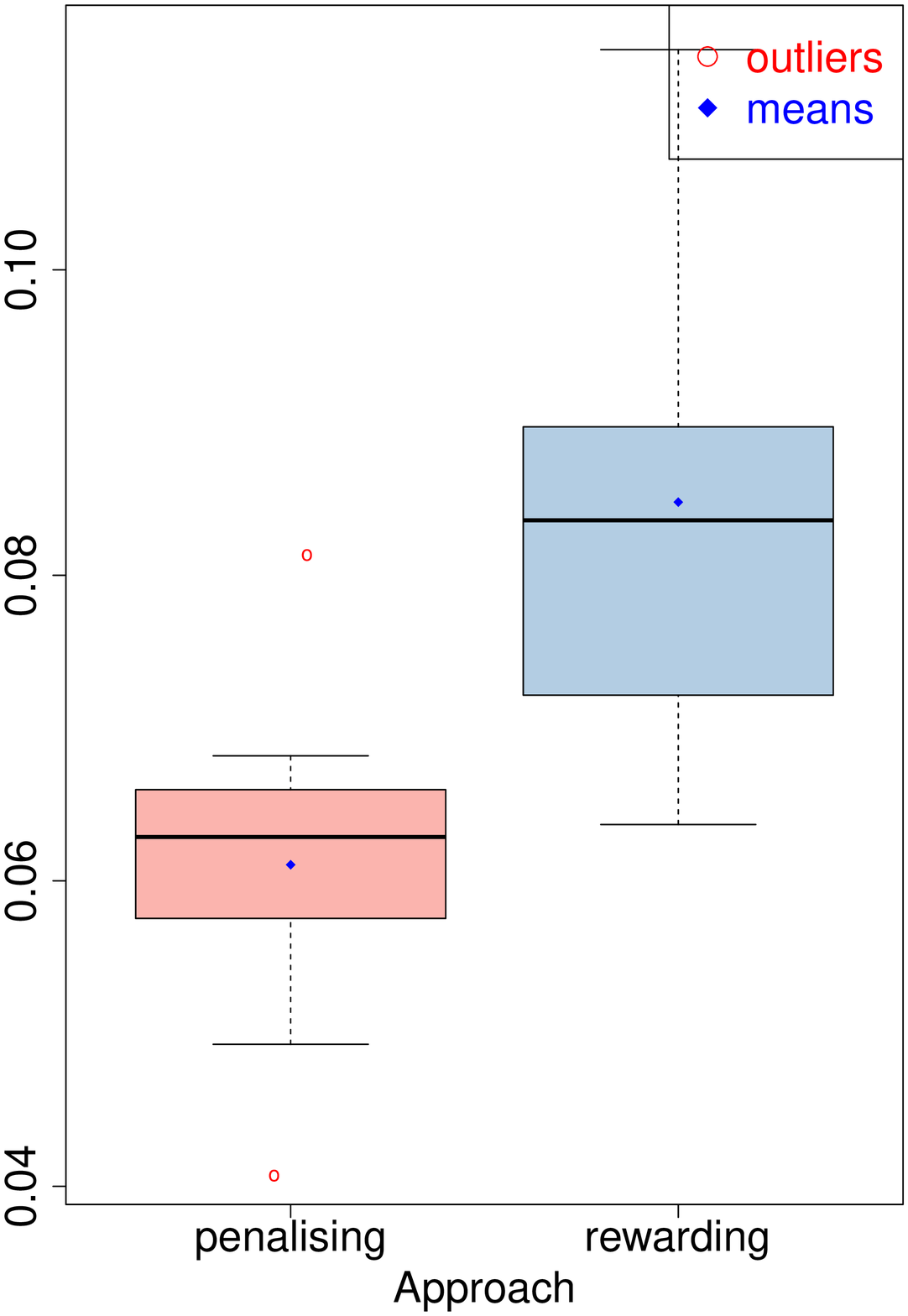}
    }%
    \subfigure[Cyclomatic complexity rate]{%
            \label{fig:boxplot_ciclomaticcomplexityrate}
            \includegraphics[width=0.2\textwidth,trim={0 0.3cm 0 0.8cm}]{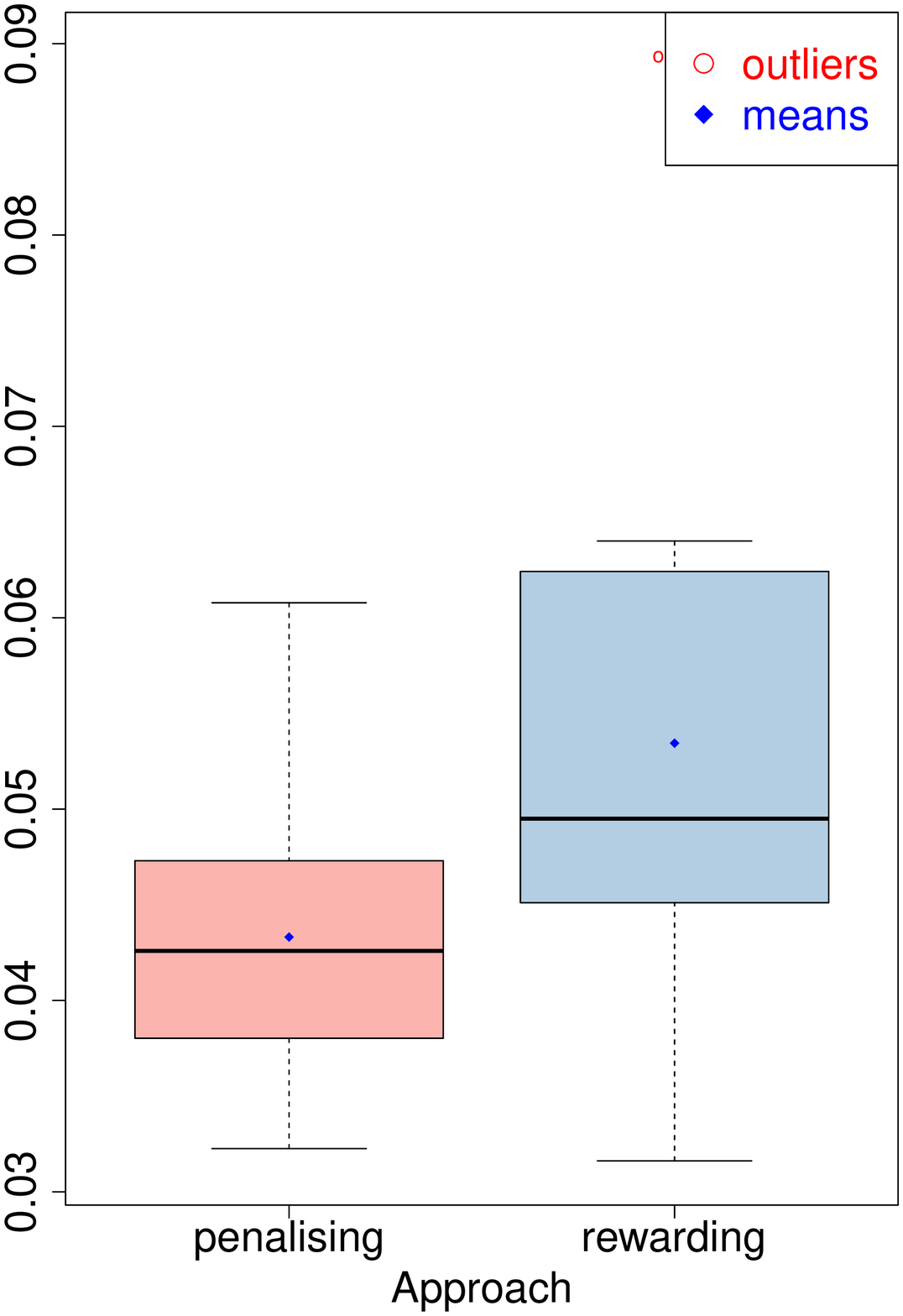}
    }%
    \caption{Metrics observation}
    \label{fig:boxplots}
    \end{center}
\end{figure*}

A descriptive analysis of the dependent variables through boxplots is carried out to compare, at first sight, if differences are appreciated. Fig.~\ref{fig:boxplots} shows the boxplots. At first glance, differences in the medians and the first quartiles can be appreciated. In some cases, in the last quartile no differences are observed, and some outliers behave differently. 

From the observation of the boxplots, it seems that the distribution of the data is not normal. Normality tests are performed by applying Shapiro-Wilk to each metric, for both strategies: penalising and rewarding.

Table~\ref{tab:normalityTestrewarding} shows the results of the Shapiro-Wilk normality tests. The null hypothesis is that the data fit a normal distribution. In almost all cases, the null hypothesis is rejected, with a $p-value < 0.05$. Non-significant p-values are highlighted in bold in Table~\ref{tab:normalityTestrewarding}. In those cases, the null hypothesis cannot be rejected.
There are 4 such cases in the group corresponding to the penalising strategy and another 3 in the group corresponding to the rewarding strategy. Therefore, in those cases, we cannot say the distribution is not normal.

Being only 7 cases out of 16, and considering the small sample size, it was decided to treat all the cases the same, assuming the data do not fit to a normal distribution and, therefore, applying a non-parametric alternative to the ANOVA or the t-test.

\begin{table}[htbp]
\caption{Shapiro-Wilk normality tests results. The $p-values > 0.05$ are marked in bold.} 
\centering
\begin{tabular}{rlrr|rr}
  \hline
  & & \multicolumn{2}{c}{Penalising} & \multicolumn{2}{c}{Rewarding} \\
 & Metric & W & p-value & W & p-value\\ 
  \hline \hline
1 & Realiability rate & 0.75 & 0.00 & 0.74 & 0.00 \\ 
  2 & Security rate & 0.78 & 0.01 & 0.54 & 0.00 \\ 
  3 & Comment density & 0.82 & 0.02 & 0.90 & \textbf{0.13} \\ 
  4 & SQALE Debt Ratio & 0.75 & 0.00 & 0.87 & 0.05 \\ 
  5 & Smells density & 0.93 & \textbf{0.38} & 0.81 & 0.01 \\ 
  6 & Duplicated lines density & 0.94 & \textbf{0.52} & 0.72 & 0.00 \\ 
  7 & Cyclomatic complexity rate & 0.95 & \textbf{0.69} & 0.94 & \textbf{0.43} \\ 
  8 & Cognitive complexity rate & 0.96 & \textbf{0.81} & 0.92 & \textbf{0.22} \\ 
   \hline
\end{tabular}

\label{tab:normalityTestrewarding}
\end{table}

\subsection{RQ1}

\begin{table*}[htbp]
\caption{Results of Asymptotic Wilcoxon-Mann-Whitney Test and Effect Size. The $p-values > 0.05$ are marked in bold.} 
\centering
\begin{tabular}{rllrrrl}
  \hline
 &        &    &         &    &            & Cohen's effect size \\  
 & Metric & $H_1$ & p-value & Z & Effect size &  classification \\ 
  \hline
1 & Reliability rate & greater & \textbf{0.23} & 0.74 & 0.15 & small effect \\ 
  2 & Security rate & greater & 0.03 & 1.93 & 0.39 & medium effect \\ 
  3 & Comment density & less & 0.03 & -1.94 & 0.40 & medium effect \\ 
  4 & SQALE Debt ratio & greater & 0.00 & 4.17 & 0.85 & large effect \\ 
  5 & Smells density & greater & 0.00 & 2.93 & 0.60 & large effect \\ 
  6 & Duplicated lines density & greater & 0.00 & 2.92 & 0.60 & large effect \\ 
  7 & Cyclomatic Complexity rate & greater & \textbf{0.96} & -1.77 & 0.36 & medium effect \\ 
  8 & Cognitive complexity rate & greater & \textbf{1.00} & -3.56 & 0.73 & large effect \\ 
   \hline
\end{tabular}
\label{tab:wilcoxresults}
\end{table*}

Table~\ref{tab:wilcoxresults} shows the results obtained from the execution of the study in R. The $p-values$ that are $> 0.05$ are marked in bold. In these three cases, the null hypothesis $H_0$ cannot be rejected. 
Hence, in the cases of the Reliability rate, the Cyclomatic Complexity rate and the Cognitive Complexity rate $H_1$ (shown in second column of Table~\ref{tab:wilcoxresults}) is not accepted.

\subsection{RQ2}

Once the Wilcoxon test for two independent samples has been used to answer RQ1, 
the effect size $r$ is calculated as the $Z$ statistic divided by the square root of the sample size ($N$) ($Z/sqrt(N)$) to answer RQ2. The $Z$ value is extracted from \verb|coin::wilcox_test()| (the case of the independent two-samples test). $N$ corresponds to the total sample size for the independent-samples test (11+13=24 in our study). The $r$ value varies from 0 to close to 1. The interpretation is explained in Section~\ref{sec:Methodology}.

Last three columns of Table~\ref{tab:wilcoxresults} show the results obtained regarding RQ2.

\section{Discussion}

The results indicate that concerning RQ1, \textbf{better results are obtained with the rewarding strategy} except in cyclomatic complexity rate and cognitive complexity rate. Regarding reliability rate, visually in the boxplot, it seems that the median and first quartiles are better. However, there is much dispersion in the last quartiles. The statistical study indicates that in this case it cannot be affirmed the result is better. 

Concerning RQ2, discarding the three previously mentioned metrics, the results show that \textbf{the improvement is significant}. The results are significant and with a large effect in three of the cases and a medium effect in the other two. The metric in which it has the highest (a very large) size effect is the Technical Debt ratio. 

Taking into account that the maximum score of the assignment is 10, the penalization strategy discounted 0.3 which represents 10\% of the 3 points assigned to the code. The other 7 points are devoted to architectural decisions, design documentation using class diagrams, sequence diagrams, statecharts, and other UML diagrams. The rewarding strategy was designed intending that a bonus of the same 0.3 discounted when penalizing can be obtained, and no more than 10\% of the maximum score can be obtained as a reward. 
In previous gamification experiences based on ranking that we have conducted, a rewarding was made by distributing points in order of ranking, but it was seen that students did not compete so much to improve as in the case the only ones who received prizes were the first three.

\section{Threats to validity}
\label{sec:threads}

\subsection{Validity of Statistical Conclusions}

As we show, this is true except in the cases of the Reliability rate, the Cyclomatic Complexity (CC) rate and the Cognitive Complexity (CogC) rate in which no statistically significant results are obtained. In all three cases it is not possible to conclude anything because the power of the test is very low, and we can neither reject nor accept the null hypothesis.

In the case of the Reliability rate, the G*Power \cite{GPower} tool was used to calculate the power of a 
Wilcoxon-Mann-Whitney test with two sample groups, one tailed (i.e., using the alternative hypothesis  ``greater''), non-normal distribution (configured as min ARE) in a \textit{Post ho}c study given the effect size $r=0.15$, $\alpha=0.05$, for a sample group 1 of size 11, and sample group of size 13. The result indicates a power ($1-\beta$) of $0.09$, which is devastating because it indicates that the probability $\beta$ of committing a II type error is about $91\%$, accepting that $H_0$ is wrong with a large probability of error. Hence, we cannot reject nor accept $H_0$.

On the other hand, G*Power tool was used to perform an \textit{a priori} analysis to compute the required sample size, given a desired $\alpha=0.05$, $\beta=0.95$ and effect size $r=0.15$. The result indicates it would require a total sample size of $2230$ projects, in two groups of $1115$ each. This differs in two orders of magnitude from the size of our sample. As a consequence, we cannot conclude anything about the Reliability rate.

The same analysis was conducted with the CC rate (CogC  rate) using G*Power. The result of the \textit{Post hoc} study indicates a power ($1-\beta$) of $0.196$ ($0.48$), which indicates that the probability of committing a II type error is about $80.4\%$ ($52\%$). Hence, we cannot reject nor accept $H_0$ in these cases either. Performing the \textit{a priori} analysis, given a desired $\alpha=0.05$, $\beta=0.95$ and effect size $r=0.36$ ($r=0.73$), a total sample size of $390$ ($96$) projects, in two groups of $195$ ($48$) each, arises. As a consequence, we cannot conclude anything about these metrics either.

\subsection{Construct validity}

Threats in this category refer to the relationship between theory and observation. In our study, this may concern mainly the selection of the metrics and the definition of the rate and density metrics. 

The same case was not used in both years to avoid the learning or copying effect. This implies a challenge when comparing the metrics that measure the projects carried out one year and the other. 
In general, metrics that express rates and not absolute values have been chosen, firstly to improve the comparison between projects carried out by different teams in the same course, but also to mitigate the fact of comparing projects carried out in two different courses to solve different cases.

The case of CC metric is special because it is really difficult to compare measures, even if they are rates of CC of two projects with different functionalities. Projects in the same course perform the same functionality, and therefore should be comparable in terms of CC. But the CC of two different projects in different courses, even defining a rate with the number of functions developed, cannot be comparable from one year to another when different functionalities are involved. In the same course, it is valid to compare the projects with each other, if they manage to provide the same functionality with a lower rate of CC. 
 
\subsection{Internal validity}

The subjects participating from one course to another are different. It could be the case that some are more prepared than others. This threat is mitigated by controlling the fact that, in both courses, they had exactly the same training during the studies and were at a time of the studies when they also had no professional experience in general. It can always happen that there are occasional cases of students with their own background, but this can happen in both courses and is diluted in the formation of the teams.

\subsection{External validity}

These mainly concern the generalisability of the reported findings. The study was conducted by obtaining measurements using SonarQube, for projects developed in Java, and in a single educational institution. In order to improve the generalisability, replications should be conducted in different educational contexts, different institutions in different countries, and with different programming languages used to develop the students' projects. 
 
In \cite{CarrotStick2020}, it is recommended that if a reward strategy is introduced in an industrial software context, a very careful design should be done, and the results of applying different reward strategies should be evaluated to ensure that they do not work to the detriment of the company. Also, some thoughts on tangible or intangible rewards are given, and on the dangers of rewarding TD elimination. A reward strategy such as the one defined in this paper, could be the basis of a proposal to be studied. The elimination of debt is not rewarded, but rather keeping TD as low as possible and a gamification strategy based on leaderboards which, in the industrial context, could be teams rather than individuals.

\section{Conclusion and future work}
\label{sec:conclusions}

In this work, we study the impact of different training strategies on
the learning outcomes related to technical debt management in
educational contexts. Penalisation and rewarding strategies are designed, applied and tested in a real environment using automatic tools. SonarQube is used to automatically obtain measures related to technical debt, and to directly implement the penalisation strategy. SonarQube is also used as the centre of the implementation of the contest (reward strategy) automated environment.
We statistically analyse and present the results of an empirical study to compare them.
The results show that in an educational context the reward strategy (the carrot), based on gamification, works better than the penalty strategy (the stick) to encourage and help the students/trainees to keep TD low and produce a high quality code. 

The implementation of the reward strategy with automatic online tools
has an added value, since it was developed in confinement during eight weeks of the academic year 2019-2020. Particularly in April, May and early June, during the first wave of the COVID-19 pandemia. We believe that it positively influenced students' motivation despite the difficulties derived from the situation.

Conducting replications of this experimental work 
in different contexts will contribute to the generalisation of the results, the acquisition of knowledge, and therefore to a better approach to the reality of the object of the study. These experiments should be enriched with interviews with the participants to understand their behaviour.

An adaptation to an industrial setting, and also to the context of open source projects development will be tackled as future work. Nevertheless, in~\cite{moldon2020HowGamificationAffectsSoftwareDevelopers}, some perverse effects of gamification elements introduced in 
collaborative development platforms, such as GitHub, are analysed. 
This sort of issues must be taken into account when adapting to that settings.

\section*{Acknowledgment}

We thanks Dr. Esperanza Manso for proofreading the paper and her valuable comments.

\bibliographystyle{plain}
\bibliography{technicaldebt2021}

\end{document}